\documentclass[journal,subeqn,caption]{IEEEtran}
\usepackage[singlelinecheck=false,labelsep=period,font=footnotesize]{caption}
\pdfoutput=1
\usepackage{mathtools}
\usepackage{tabularx}
\usepackage{cite}
\usepackage{float}
\usepackage{multicol}
\usepackage{multirow}
\usepackage{hyperref}
\usepackage{graphicx}
\usepackage{amsfonts}
\usepackage{amsmath}
\usepackage{color}
\usepackage{epstopdf}
\usepackage{caption}
\usepackage{subcaption}
\usepackage{mathtools, nccmath}
\usepackage{multirow}
\usepackage{bbm}
\usepackage{marginnote}
\usepackage{tikz}
\usepackage{enumitem}
\usepackage{dblfloatfix}
\usepackage{comment}
\usepackage{eurosym}
\usepackage{makecell}
\usepackage[switch]{lineno}
\usetikzlibrary{arrows,decorations.pathmorphing,backgrounds,fit,positioning,shapes.symbols,chains}
\usepackage{colortbl}

 \ifCLASSINFOpdf
\else
\fi
\begin{document}
\title{EV-based Smart E-mobility System -- Part I: Concept}
\author{I. Pavi\'c, \textit{Student Member}, \textit{IEEE}, 
\thanks{Ivan Pavi\'c, Hrvoje Pand\v{z}i\'c and Tomislav Capuder are with the University of Zagreb Faculty of Electrical Engineering and Computing, Zagreb HR-10000, Croatia (e-mails: \href{mailto:ivan.pavic@fer.hr}{ivan.pavic@fer.hr},  \href{mailto:hrvoje.pandzic@fer.hr}{hrvoje.pandzic@fer.hr}, \href{mailto:tomislav.capuder@fer.hr}{tomislav.capuder@fer.hr}).}% Their work has been supported in part by Croatian Science Foundation and Croatian TSO (HOPS) under the project Smart Integration of RENewables - SIREN (I-2583-2015) and Croatian Environmental Protection and Energy Efficiency Fund, as a part of the H2020 ERA-Net Smart Grids Plus funding scheme, under the project microGRId Positioning - uGRIP.}
H. Pand\v{z}i\'c, \textit{Senior Member}, \textit{IEEE},
T. Capuder, \textit{Member}, \textit{IEEE}
}
%
%
% make the title area
\maketitle

\begin{abstract}
The first of this two-paper series proposes and elaborates a concept of electric vehicle (EV)-based e-mobility system. To this end, models designed to reap the benefits of EVs' flexibility in the literature almost exclusively consider charging stations as active players exploiting the EVs' flexibility. Such stations are seen as static loads able to provide flexibility only when EVs are connected to them. However, this standpoint suffers from two major issues. First, the charging stations need to anticipate some important parameters of the incoming EVs, e.g. time of arrival and departure, state-of-energy of the EV's battery at arrival and the required state-of-energy at its departure. Second, it observes the EVs only when they are connected to these charging stations, thus overlooking the arbitrage and charging opportunities when the EVs are connected to other charging stations.

This paper proposes a new viewpoint, where EVs are observed as dynamic movable storage systems which can provide flexibility at any charging station. The paper defines both systems, the existing one, where the flexibility is viewed from the standpoint of charging stations, and the proposed one, where the flexibility is viewed from the EVs' standpoint. %list their up/downsides and discuss their implementation issues. The optimization model of the two systems will be designed which will provide a proof that EVs offer more flexibility in a proposed EV-based concept then in the commonly used CS-based.
\end{abstract}
\begin{IEEEkeywords}
Electric vehicles, Charging stations, Aggregator, Electricity market.
\end{IEEEkeywords}

\section{Introduction}
Increasing shares of renewable energy sources (RES) entail lower marginal price of energy, but at the same time decrease system controllability. To allocate sufficient flexibility, new technologies (e.g. energy storage) or new concepts (e.g. demand response) are introduced. Electrification of the transport sector provides an attractive new flexibility source. If electric vehicles (EVs) are charged uncontrollably \cite{Muratori2018}, i.e. charging at maximum power until fully charged, power system's hunger for flexibility increases, calling for additional investments in peaking units and grid infrastructure upgrades. On the other hand, if EVs are charged in a controllable manner \cite{Wolinetz2018}, they resemble features of both demand response and energy storage. Shifting their charging times represents the aspect of demand response. This is often referred to as Grid-to-Vehicle (G2V) mode, which requires unidirectional controllable chargers \cite{Xu2018}. A possibility to discharge a part of the surplus energy when not needed for motion, often referred to as Vehicle-to-Grid (V2G) mode, corresponds to the energy storage aspect of EVs and requires bidirectional controllable chargers \cite{Ivan1}, \cite{Ivan2}. Detailed overviews of EV charging modes are available in \cite{Liu2013} and \cite{Monteiro2016}.

This paper proposes a new concept of using the EV flexibility more efficiently in a world with a multitude of new data streams relying on information-communication technologies in vehicles and without any loss of comfort for EV drivers. We argue that the state-of-the-art literature, research projects and e-mobility sector conceive the smart e-mobility in a way which leads to an underutilization of EV flexibility and to insufficient financial returns. The usual understanding of smart e-mobility is that Charging Stations (CSs) use EVs to provide flexibility to the power system (CS-based concept), whereas this paper challenges this concept and reverses the roles by stating that EVs themselves are actually smart players providing flexibility and the CSs are merely an enabling infrastructure (EV-based concept). 

The \emph{smart e-mobility} term used in this paper refers to an advanced multisector system where the main actors are: EVs, CSs, Electric Vehicle Aggregators (EVAs), power grid and electricity market operators. Merchant actors within this ecosystem have at their disposal smart EV charging and discharging to provide flexibility to the power system and in return receive monetary reward. This paper analyzes a basic illustrative example with 3 EVs and 3 CSs. The purpose of the example is to highlight certain issues in the state-of-the-art, after which, in the second part of this two-paper series, we define the proposed EVA mathematical model and demonstrate how it eliminates the issues of the current state-of-the-art through a detailed case study.

Contribution of this two-part paper series consists of the following:

\renewcommand{\labelenumi}{{\roman{enumi})}}
\begin{enumerate}
%\item defining two different smart e-mobility concepts, CS-based and EV-based, and analyzing their pros and cons in real life,
%\item to steer the research and industry toward the EV-based concept as it yields better results for EV users and power system,
\item design of a novel EVA model tracking the EVs at all locations thus capturing all relevant battery information,
\item formulation of the proposed EV-based model,% provides improved results for both the EV users and the society in general as compared to the conventional CS-based models,
\item systematic and rigorous comparative assessment
of the CS-based and EV-based models,
\item demonstration that models without relevant features, such as power levels and grid tariffs, drive the obtained results in a wrong direction.
\end{enumerate}

%The main contribution of the paper is definition of two different smart e-mobility concepts, CS-based and EV-based, illustrating their main features and analyzing their pros and cons. The conclusions could steer the research and industry toward the EV-based concept as it yields better results for EV users and the power system as a whole.

\section{Illustrative Example} \label{sec:example}
\subsection{Assumptions and Description}
\begin{figure}[t]
    \centering
    \includegraphics[trim={0.1cm 0.6cm 0.01cm 0.1cm},clip,width=\linewidth]{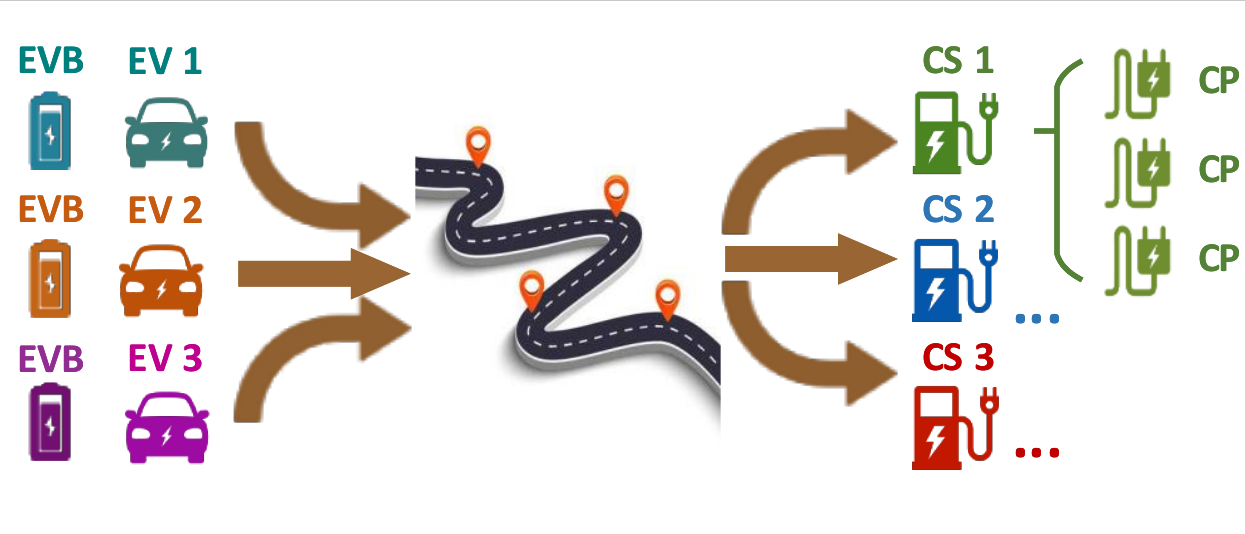}
    \caption{Illustrative example - overview}
    \label{fig:main}
\end{figure}

\begin{figure}[t]
    \centering
    \includegraphics[trim={0.5cm 0.1cm 0.1cm 0.1cm},clip,width=\linewidth]{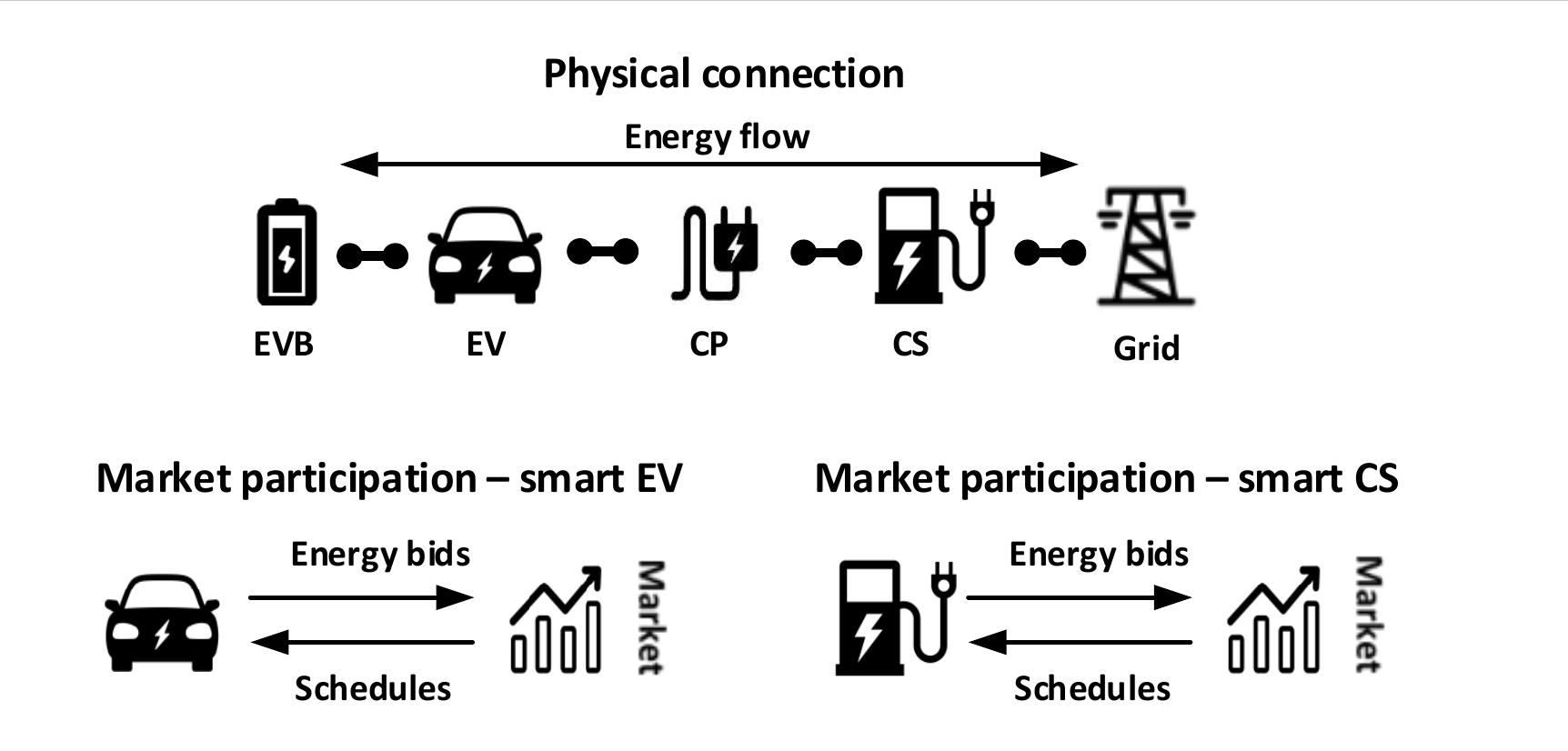}
    \caption{Physical connection and market participation of EV-based and CS-based smart e-mobility models}
    \label{fig:conn}
\end{figure}

An illustrative example presented in Figures \ref{fig:main}, \ref{fig:conn}, and \ref{fig:graph} compares the current smart e-mobility CS-based model with the proposed EV-based concept. Several simplifications and assumptions are made to keep this example concise. We observe three EVs and three CSs (Figure \ref{fig:main}) and their behavior through a 24-hour period with 1-hour time resolution. Each EV can be charged at different CSs and each driving period, i.e. period when EV is not connected to any CS, lasts one hour. Each EV has one Battery (EVB) and one On-Board Charger (OBC), while each CS encompasses three Charging Points (CPs), meaning it can serve all three EVs at a time. %They could be understood as a standalone chargers with three plugs. 
All three CPs within a CS are AC and have chargers of same power capacity. %Each EV can be parked at one CS at a time.

Let us assume that both EVs and CSs can individually participate in the wholesale electricity market\footnote{Currently this is done through aggregators due to energy bid thresholds in most markets.}, namely the day-ahead market, and that their objective is to minimize the purchasing costs of electricity for mobility purposes and/or to maximize earnings through energy arbitrage. Other types of flexibility revenues (such as balancing services) are neglected for the sake of brevity. 

A smart e-mobility system can therefore be conceived as an EV-based or a CS-based, as illustrated in Figure \ref{fig:conn}. In the former model, the EVs are the smart entities negotiating market strategy while the CSs are merely an infrastructure with their technical constraints (CP power capacity) and economic parameters (CS utilization fee). The latter model observes the same entities, but from an opposite standpoint. The CSs are the smart entities negotiating market strategy, while the EVs only impose technical constraints (OBC power capacity) and economic charges (battery utilization fee). The CSs must pay a fee to use the EVs' physical equipment (battery) and the energy stored within the EVBs when performing arbitrage (V2G mode). On the other hand, they receive payments by the EVs for the energy they charge for driving purposes. 

\subsection{EV-based vs. CS-based Smart E-mobility Model}
\begin{figure*}[t]
    \centering
    \includegraphics[width=16cm]{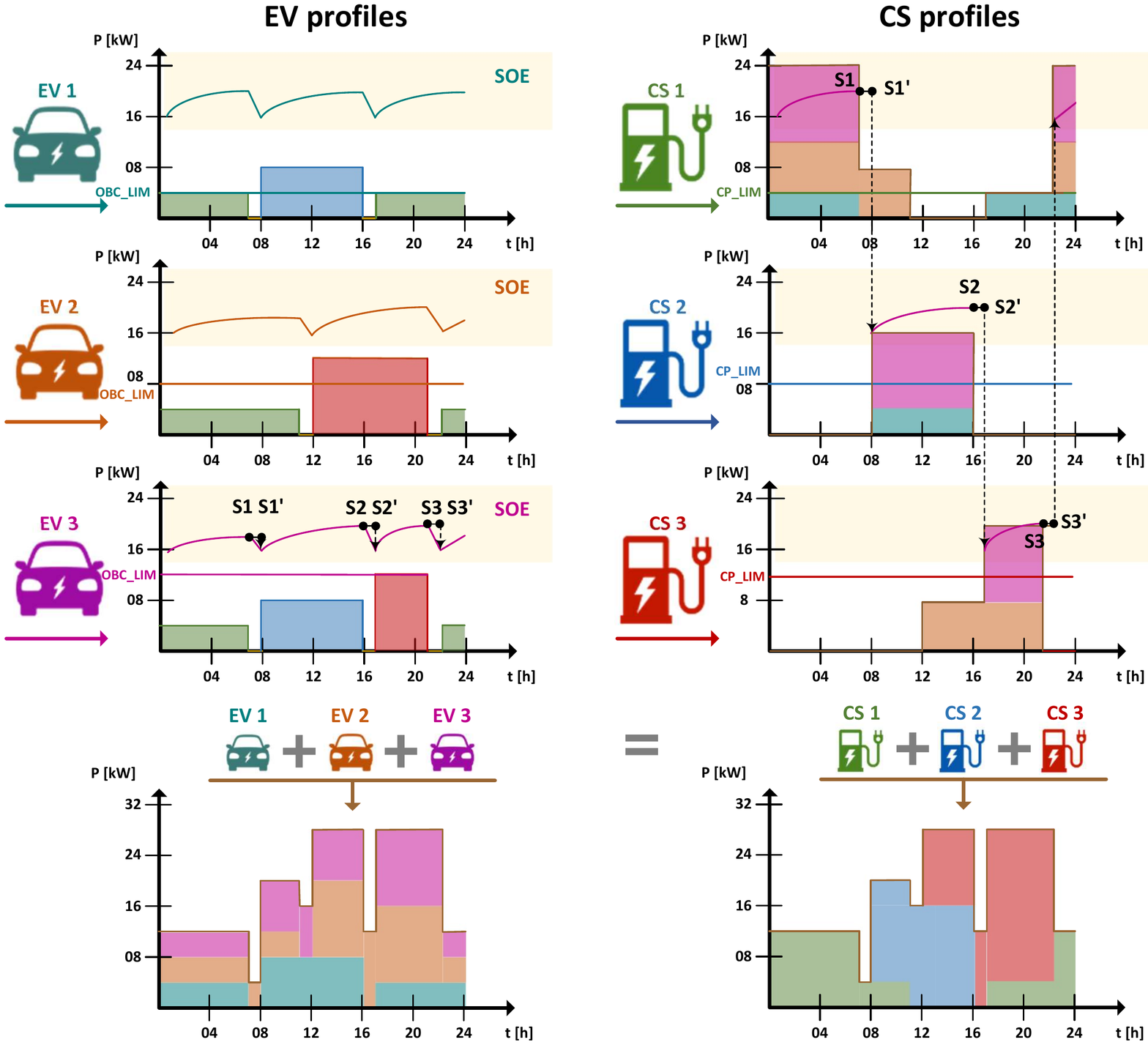}
    \caption{Illustrative example - daily curves for three EVs and three CSs}
    \label{fig:graph}
\end{figure*}

We use the graphs in Figure \ref{fig:graph} to describe the differences between the EV-based and the CS-based smart e-mobility models. The graphs to the left show charging profiles of the three EVs, while the ones to the right show charging profiles of the three CSs. All graphs are created from the same data, but observed from different viewpoints: graphs to the left are relevant for the EV-based, while the ones to the right are relevant for the CS-based smart e-mobility model.

EVs in Figure \ref{fig:graph} are shown in different colors: EV1 -- turquoise, EV2 -- orange, and EV3 -- purple. Their respective OBC maximum powers (OBC\_LIM) are marked with straight lines: EV1 -- low-power OBC (4 kW), EV2 -- medium-power OBC (8 kW), and EV3 -- high-power OBC (12 kW). The EVs can charge at three CS types: CS1 is a home charger (4 kW) -- green, CS2 is charger at work (8 kW) -- blue, and CS3 is charger at a shopping mall (12 kW) -- red. EVs have different driving profiles. EV1 has a \emph{home-work-home} profile: it is connected to CS1 from midnight to 07:00, drives to CS2 where it stays from 08:00 to 16:00, and drives back to CS1, where it is connected from 17:00 to midnight. Charging profile of EV2 is \emph{home-mall-home}, while EV3's profile is \emph{home-work-mall-home}. Each charging period is colored according to the corresponding CS.

The CS charging curves are composed of the charging curves of the EVs connected to it. For example, the graph for CS1 (upper-right in Figure \ref{fig:graph}) shows that all three EVs are connected to it from 00:00 to 07:00. The power required at CS1 during that time is the sum of the OBC powers of the EVs using it. EV2 (orange) is staying longer at CS1 (until 11:00), while EV1 (turquoise) comes back home earlier than others (at 17:00). Since EV1 and EV3 are at work during the morning and midday hours, CS2 has two connected EVs from 08:00 to 16:00 (turquoise EV1 and purple EV3), and no connected EVs in other hours. EV2 goes to a shopping mall, where it charges at CS3 from 12:00 to 21:00, while EV3 goes to the mall after work form 17:00 to 21:00 (third graph on the right-hand side). The areas in the graphs to the right correspond to stacked OBC powers, while the maximum CP power limits are indicated with fixed straight lines. In instances where \emph{CP\_LIM} is lower than \emph{OBC\_LIM}, the CP is the limiting factor for charging power.

The lower graphs in Figure \ref{fig:graph} are aggregate curves based on the EVs' behavior (left) and the CSs' behavior (right). The colors display which EV (left) or CS (right) contributes to the aggregated behavior at a specific period of time. The outline curve is the same in the left and the right graphs, meaning that if there is only one central aggregation entity that oversees all EVs and CSs, it does not matter whether it is defined as an EV- or a CS-based. However, it does matter when multiple aggregators enter the market.

The areas with the yellow background in the graphs in Figure \ref{fig:graph} show the EV state of energy (SOE) throughout the day. In graphs to the left, each SOE curve corresponds to the corresponding EV, while in graphs to the right only the SOE curve of EV3 is displayed for simplicity.

\subsection{Data Transfer} \label{sec:data}

Different data forms must be exchanged between the EVs and the CSs which is essential for correct smart e-mobility operation in both the EV- and CS-based system. In the EV-based system the CS data must be sent to EVs, while in the CS-based system the EV data must be sent to CSs.

Required EV data are:
\renewcommand{\labelenumi}{{\roman{enumi})}}
\begin{enumerate}
\item technical data -- parameters such as OBC power levels, battery capacity, etc.,
\item infrastructure cost -- expenses arising from EV usage apart from mobility reasons, such as V2G battery degradation,
\item preferences -- EV users' desires related to charging, such as minimum SOE under which an EV does not offer flexibility, targeted SOE at some point in time, etc.,
\item behaviour -- historic driving/parking data which serve as a base for future EV behaviour forecasts.
\end{enumerate}

Required CS data are:
\renewcommand{\labelenumi}{{\roman{enumi})}}
\begin{enumerate}
\item technical data -- parameters such as CP connector type and CP power levels,
\item infrastructure cost -- expenses arising from CS usage for any kind of charging and discharging, e.g. CS operation and maintenance cost, CS investment return, and grid fees.
\end{enumerate}

\section{Issues and Proposed Solution} \label{sec:issues}

In the CS-based smart e-mobility system the CSs submit their individual bids in the market. Each of them runs their own optimization algorithm based on their own predictions. However, this results in the issues individually elaborated below. 

\subsection{Issue 1 -- Insufficient Information on EVs' Behavior at Other CSs}

\subsubsection{CS-based Issue}
\hspace*{\fill}

The first issue is that a CS only tracks the EVs' SOE in the periods when they are connected to it. From the mathematical standpoint, power to be charged/discharged and the SOE while the EVs are either parked at other premises or driving are unknown and included in the model as stochastic parameters. Only when EVs are connected to this CS those values become controllable variables. If observing the SOE curve of EV3 in Figure \ref{fig:graph}, it is broken down into several segments (at points S1-S3), where each CS can see only one part of it but not the entire daily curve. This is a major drawback since the values of the (dis)charging variables should come directly from forecasting the four main attributes of each EV:

\renewcommand{\labelenumi}{{\roman{enumi})}}
\begin{enumerate}
\item arrival time of vehicle $v$ ($t^\mathrm{ARR}_v$),
\item SOE at arrival ($SOE^\mathrm{ARR}_{v}$),
\item departure time ($t^\mathrm{DEP}_v$),
\item required SOE at departure ($SOE^\mathrm{DEP}_{v}$).
\end{enumerate}

For the CSs in the presented example, the following stands for EV3:
\begin{itemize}
    \item CS1 forecasts $t^\mathrm{DEP}$ and $SOE^\mathrm{DEP}_{v,cp1}$ at S1 and $t^\mathrm{ARR}$ and $SOE^\mathrm{ARR}_{v,cp1}$ at S3’,
    \item CS2 forecasts $t^\mathrm{ARR}$ and $SOE^\mathrm{ARR}_{v,cp2}$ at S1’ and $t^\mathrm{DEP}$ and $SOE^\mathrm{DEP}_{v,cp2}$ at S2,
    \item CS3 forecasts $t^\mathrm{ARR}$ and $SOE^\mathrm{ARR}_{v,cp3}$ at S2’ and $t^\mathrm{DEP}$ and $SOE^\mathrm{DEP}_{v,cp3}$ at S3.
\end{itemize} 

The CSs must do the same for all EVs coming to charge. Mathematically, this is represented as follows:
\begin{flalign}
%\mathrm{if} \quad t\in \Omega^{T_\mathrm{parked\_at\_observed\_CS}}_{v,cp}\nonumber
%\end{flalign}
%\begin{gather}
\mathrm{if} \quad& t\in \Omega^{T_\mathrm{parked\_at\_observed\_CS}}_{v,cp}
\nonumber\\
&soe^{\text{EV}}_{v,t}\! =\! soe^{\text{EV}}_{v,t-1} \!+\! e^{\text{SCH}}_{v,t} \!\cdot\! \eta^{\text{SCH}}
\!-\! e^{\text{DCH}}_{v,t} \!/\! \eta^{\text{DCH}}; \label{e_SOE4V}
\\
\mathrm{else - if} \quad& t=t^\text{ARR}_{v,cp}
\nonumber\\
&soe^{\text{EV}}_{v,t} = SOE^\text{ARR}_{v,cp}; \label{SOE_ARR}
\\
\mathrm{else - if} \quad& t=t^\text{DEP}_{v,cp}
\nonumber \\
&soe^{\text{EV}}_{v,t} \geq SOE^\text{DEP}_{v,cp};
\label{SOE_DEP}
\\
\mathrm{else}\quad &t\in \Omega^{T_\mathrm{driving\_or\_parked\_at\_other\_CS}}_{v,cp}
\nonumber \\
&soe^{\text{EV}}_{v,t} \quad \text{unconstrained} \quad \forall v,t.
\label{SOE_other}
\end{flalign}

The first equation tracks an EVB during the period when the EV is parked at the observed CS, with variables $soe^{\text{EV}}_{v,t}$, $e^{\text{SCH}}_{v,t}$ and $e^{\text{DCH}}_{v,t}$ denoting the EV's SOE, energy charged and discharged, respectively, and $\eta^{\text{SCH}}$ and $\eta^{\text{DCH}}$ the corresponding efficiencies. The second and the third constraints set the $soe^{\text{EV}}_{v,t}$ at arrival/departure based on the SOE forecasts or requirements. The periods when an EV is driving or parked at other CSs are not explicitly modeled and its behavior during these periods can only be considered through the forecasted values of unknown parameters. 

The questions that inspired this research were: How would each of the CSs forecast the four uncertain values (arrival time, SOE at arrival, departure time and required SOE at departure) for all the EVs with sufficient accuracy? How would they anticipate the EVs' behavior while driving and especially while at other CSs? One option is that each EV sends its data to all the CSs where it could potentially park and charge. Another option is that each CS sends its own forecasts for each EV to all CSs in surroundings, i.e. all the CSs should optimize their portfolio in a joint optimization or using separate optimizations with coupling SOE constraints. On top of the issue of global optimality of such approach, the amount of data to be transmitted becomes critical and data security issues could easily render such model inapplicable.

\subsubsection{EV-based Solution}
\hspace*{\fill}

In the EV-based smart e-mobility system, the three EVs in Figure \ref{fig:graph} submit their individual bids to the market operator. Each of them runs its own independent optimization algorithm based on own predictions. Contrary to the CS-based system, each EV knows its behavior (SOE curve) throughout the day wherever it is. From the mathematical standpoint, power to be (dis)charged and the SOE is always known to the EV. If the SOE curve of EV3 in Figure \ref{fig:graph} is observed, EV3 sees it as a continuous line without interruptions at points S1-S3, while the CSs see only their portion of this curve. The EV-based model can thus be mathematically represented as follows:

\begin{gather}
soe^{\text{EV}}_{v,t} = soe^{\text{EV}}_{v,t-1} + e^{\text{SCH}}_{v,t} \cdot \eta^{\text{SCH}} 
- e^{\text{DCH}}_{v,t} / \eta^{\text{DCH}} 
\nonumber\\
- E^{\text{RUN}}_{v,t} / \eta^{\text{RUN}} 
+ e^{\text{FCH}}_{v,t} \cdot \eta^{\text{FCH}}
 \quad \forall v,t;
 \label{SOE_c}
\end{gather}
It sets the EVs' SOE considering the SOE from the previous time step, charging at a slow CS (SCH), energy discharged in V2G mode (DCH), discharged for driving purposes (RUN), and energy charged at fast charging stations (FCH). Compared to equations (1)--(4) in the CS-based system, this model observes and controls all variables at all time periods. The forecasting effort is drastically reduced and simplified since the EV predicts its own behavior, while in the CS-based system each CS must predict behavior of a multitude of EVs. There is no need for the EV-to-CS communication nor for additional CS-to-CS communication. Each EV keeps its driving/parking information and its technical data to itself and does not send any data to other entities. The complexity of data flow is reduced, while its security is increased as compared to the CS-based model.

\subsection{Issue 2 -- Inability to Transfer Flexibility between CSs}

\subsubsection{CS-based Issue}
%\hspace*{\fill}

The second issue in the CS-based system relates to daily human activities and the way the CSs are usually organized. In our example, CS1 is a home charger and has access to the EVs mostly during the night. On the other hand, CS2 has EVs connected to it only during daytime, while the EVs are at CS3 mostly during the evening periods. When performing energy arbitrage, the energy should be shifted from peak- to low-price periods. Usually, the prices are lower during the night (when the consumption is low) and midday (when PV generation is high and load is at its local minimum), while the peak prices occur in the morning and evening (when PV generation is low and consumption high). The CSs aiming to perform energy arbitrage with EVs should thus roughly follow the sequence: night$\rightarrow$charging, morning$\rightarrow$discharging, midday$\rightarrow$charging, evening$\rightarrow$discharging. CS1 has only one EV connected to it in the morning and the evening so it cannot discharge all the EVs at peak periods. At midday it does not have any EVs connected to it and thus cannot recharge them. CS2 cannot transfer energy from night to evening periods because it does not have any EVs connected to it in the evening, but can discharge the EVs in the morning and recharge them at midday. However, to have enough energy to discharge EVs in the morning it must communicate with CS1 and request additional charging (more than necessary for mobility). CS3 can discharge EVs in the evening, but it needs to communicate the additional energy with CS2.  

\subsubsection{EV-based Solution}
\hspace*{\fill}

The EV-based concept follows the EVs throughout the day. If behavior of EV3 in Figure \ref{fig:graph} is observed, it can provide the optimal charge--discharge sequence following the typical daily price curve elaborated above. It can charge during the night at CS1 and discharge in the morning at CS2, where it can also recharge around midday. Then, it can discharge in the evening at CS3 and start charging at CS1 late in the evening. In the EV-based system, the EV flexibility can thus be fully exploited without the need for CS-to-CS communication. To summarize, the proposed EV-based concept results in higher savings, no privacy issues and lower communication burden.

\subsection{Issue 3 -- Insufficient Power Constraints} \label{sec:issue3}
\begin{table*}[!b]
\centering
\caption{Categorization of research papers related to \emph{Issues 1--4} (comm. -- commercial; ch. -- charging; inf. -- infrastructure; deg. -- degradation)}
\label{tab:issues}
 \begin{tabular}{|c|c c c|c c c|} 
 \hline
  & \multicolumn{3}{|c|}{\emph{Issue 1} -- insufficient information on EVs’ behavior at other CSs} & \multicolumn{3}{|c|}{\emph{Issue 2} -- inability to transfer flexibility between CSs}\\
 \hline
\makecell{Literature \\ type} & \makecell{CS-based} & \makecell{ EV-based \\ with only 1 CS} & EV-based & Households & \makecell{Work/comm. \\ ch. station } & Multiple\\  
 \hline\hline
 \makecell{Smart homes/ \\ microgrids} & \makecell{\cite{Erdinc2015}, \cite{Wi2013}, \cite{Pal2018}, \\ \cite{Rana2018},  \cite{Naghibi2018}, \\ \cite{Tushar2014}, \cite{Paterakis2016}} & \cite{Melhem2017}, \cite{Kikusato2018}, \cite{Zhang2016comm} & - &
  \makecell{\cite{Erdinc2015}, \cite{Wi2013}, \cite{Pal2018},\\  \cite{Rana2018}, \cite{Naghibi2018}, \cite{Tushar2014},\\  \cite{Paterakis2016}, \cite{Melhem2017}, \cite{Kikusato2018}}&  - & \cite{Zhang2016comm}
 \\ 
 \hline
  \makecell{EV \\ aggregators} & \makecell{\cite{Nguyen2014}, \cite{Wei2016}, \cite{Vagropoulos2015}, \\ \cite{Bessa2013}, \cite{Wu2016}, \cite{Bessa2012}, 
  \\ \cite{Kaur2019}, \cite{Goebel2016}, \cite{Vagropoulos2013} } & \cite{Sarker2016}, \cite{Ortega-Vazquez2013} & - &
  \makecell{\cite{Nguyen2014}, \cite{Wei2016}, \cite{Vagropoulos2015}, \\ \cite{Bessa2013}, \cite{Wu2016}, \cite{Kaur2019}, \\ \cite{Vagropoulos2013}, \cite{Sarker2016}, \cite{Ortega-Vazquez2013},}&  - & \cite{Bessa2012}, \cite{Goebel2016}
 \\ 
 \hline
  \makecell{Parking lots/ \\ ch. stations} & \makecell{\cite{Zhang2018}, \cite{Zhang2016}, \cite{Zhang2017},  \cite{Awad2017},  \\
   \cite{Kuran2015}, \cite{Akhavan-Rezai2016}, \cite{Mehrabi2018}, \\ \cite{Zheng2019}, \cite{Zhou2019}, \cite{You2016}}  & - & - &
    \cite{Mehrabi2018} & \makecell{ \cite{Zhang2018}, \cite{Zhang2016}, \cite{Zhang2017},\\ \cite{Awad2017}, \cite{Kuran2015},   \cite{Akhavan-Rezai2016}, \\ \cite{Zheng2019}, \cite{Zhou2019}, \cite{You2016}} & - 
   \\
 \hline
  \makecell{Proposed \\ concept} & - & - & \checkmark &
  -&-& \checkmark
  \\
  \hline\hline
  
  & \multicolumn{3}{|c|}{\emph{Issue 3} -- insufficient power constraints} & \multicolumn{3}{|c|}{\emph{Issue 4} -- incomplete costs}\\
 \hline
\makecell{Literature \\ type} & 
Fixed & \makecell{CP or \\ OBC only} & \makecell{Both CP \\ and OBC} & 
\makecell{No grid/\\inf./deg. cost } & \makecell{With grid fee/ \\ inf. cost} & \makecell{With \\ deg. cost}
\\  
 \hline\hline
 \makecell{Smart homes/ \\ microgrids} &
 \makecell{\cite{Erdinc2015}, \cite{Wi2013}, \cite{Pal2018},  \\    \cite{Rana2018},\cite{Naghibi2018}, \cite{Tushar2014}, \\ \cite{Melhem2017}, \cite{Kikusato2018}, \cite{Zhang2016comm}} & \cite{Paterakis2016} & - &
  \makecell{\cite{Erdinc2015}, \cite{Wi2013}, \cite{Rana2018},\\  \cite{Naghibi2018}, \cite{Tushar2014}, \cite{Paterakis2016}, \\  \cite{Kikusato2018}, \cite{Zhang2016comm} } & \cite{Pal2018}  & \cite{Melhem2017}
 \\ 
 \hline
  \makecell{EV \\ aggregators} &
  \makecell{\cite{Nguyen2014},  \cite{Vagropoulos2015}, \cite{Bessa2013}, \\ \cite{Wu2016}, \cite{Kaur2019}, \cite{Goebel2016}, \\  \cite{Vagropoulos2013}, \cite{Sarker2016}, \cite{Ortega-Vazquez2013}} & \cite{Wei2016}, \cite{Bessa2012} & - & 
   \makecell{\cite{Nguyen2014}, \cite{Wei2016},  \cite{Vagropoulos2015}, \\ \cite{Bessa2013},  \cite{Wu2016}, \cite{Bessa2012},\\   \cite{Goebel2016}, \cite{Vagropoulos2013}, \cite{Ortega-Vazquez2013} } & - & \cite{Kaur2019}, \cite{Sarker2016}
 \\ 
 \hline
  \makecell{Parking lots/ \\ ch. stations} &
   \makecell{\cite{Zhang2018}, \cite{Zhang2016},  \cite{Zhang2017}, \\ \cite{Kuran2015},  \cite{Akhavan-Rezai2016}, \cite{Mehrabi2018}, \\ \cite{Zheng2019}, \cite{Zhou2019}, \cite{You2016} } & - & \cite{Awad2017} &
    \makecell{\cite{Zhang2018}, \cite{Zhang2016},  \cite{Zhang2017}, \\ \cite{Kuran2015},  \cite{Akhavan-Rezai2016}, \cite{Mehrabi2018} , \\ \cite{Zheng2019}, \cite{Zhou2019}} & \cite{Awad2017} & \cite{You2016} 
   \\
 \hline
  \makecell{Proposed \\ concept} & 
  -&-& \checkmark &
  -& \checkmark/\checkmark & \checkmark
  \\
  \hline
\end{tabular}
\end{table*}

\subsubsection{CS-based Issue}
\hspace*{\fill}

Throughout the day, EVs with different OBC power capacities park at CSs with various power capacities. This issue is illustrated in the graphs to the right in Figure \ref{fig:graph}, where each CP installed power capacity is shown with a fixed value, CP\_LIM, while the EVs' OBC power constraints are shown as stacked colored areas. If the OBC power constraints are omitted, the CSs could end up scheduling higher power than technically possible to deliver, e.g. EV1 at CS2. On the other hand, if the OBC power constraint is higher than the CP power constraint, the CP constraint is binding and does not affect the EV scheduling, e.g. EV2 at CS1.  

Such events can cause differences between the scheduled and delivered energy and lead to additional balancing costs. The OBC installed power is an additional parameter that all EVs must communicate to the CSs or CSs must anticipate, which can lead to errors. Furthermore, this EV-to-CS communication is highly inconvenient due to large amount of dynamic data as well as security issues. 

\subsubsection{EV-based Solution}
\hspace*{\fill}

EVs change their location during the day. In our example EV1 and EV2 park at two, while EV3 parks at three different locations. Since CSs have different installed charging powers, the EVs must anticipate the installed power of the CSs where they park. %However, this is not a problem since the CS powers are public. %to correctly anticipate their charging/discharging on market$ 
This is illustrated in Figure \ref{fig:graph} on graphs to the left, where each EV’s OBC installed power (OBC\_LIM) is shown as a fixed value, while the CSs' capacities vary through the day (visualized as stacked colored areas). If the CS power constraint is omitted, the EVs whose OBC is of higher power than the CS's maximum power could schedule more charging power than possible in reality, e.g. EV2 during night/morning parked at CS1. 

As in the CS-based concept, both the OBC and CS maximum charging power constraints need to be included in the optimization model. However, most CSs publicly publish their chargers' technical parameters, such as connector type and installed power, and EVs can easily download the required data. The EV-to-CS communication is again avoided making the EV-based system easier to implement than the CS-based system.

\subsection{Issue 4 -- Incomplete Costs}

\subsubsection{CS-based Issue}
\hspace*{\fill}

Each EV must pay energy cost for its basic mobility charging in the electricity market (through its CS supplier). Apart from energy expenditures, each load must pay a grid fee (upper part of Figure \ref{fig:conn}). At lower voltages, loads generally pay higher fees since they use grid of several voltage levels. CSs are connected to the low voltage distribution grid and the grid fees account for a significant share in their total costs transferred to the EVs. To properly address the cost of EV charging, grid fees  must be taken into account. \footnote{Generation facilities mostly do not pay the grid fees. In case of V2G discharging, such fees could have a major effect on its financial profitability.}

Apart from energy cost and grid fees, there is a cost associated with remuneration between EVs and CSs. When it comes to basic mobility charging, EVs pay fees to the CSs to recover the operation and maintenance costs, as well as the investment. However, when CSs use EVBs for energy arbitrage or other actions beside basic mobility charging, they should pay a fee to the EVs for using their battery since increased battery cycling causes it to degrade faster. In the CS-based system, a CS must obtain data from EVs on their infrastructure (battery) costs. Again, the EVs must send their private data to all relevant CSs. 

\subsubsection{EV-based Solution} \hspace*{\fill}

In the EV-based system, EVs obtain data on CS infrastructure costs and grid fees. Unlike the EVs, the CSs are public and already publish their prices online to attract EVs. In the proposed EV-based system, EVs must pay a fee to CSs whenever they use them for energy arbitrage and/or basic mobility charging. The EV-to-CS communication is not necessary as the relevant CS data are available online. 

\section{Current State-of-the-art, Industry Practices and Proposed Concept}

\subsection{Literature Review} \label{sec:lit}

State-of-the-art literature on smart e-mobility scheduling can be divided into several research approaches. Table \ref{tab:issues} summarizes the literature considering three topics (smart home/microgrids, EV aggregators, smart parking lots/charging stations) and the way they tackle the four issues detected in Section \ref{sec:issues}. Under \textit{Issue 1} we add an intermediate step between the CS-based and EV-based concepts for papers using equations similar to \eqref{SOE_c}, but not specifying chargers or considering only residential chargers. 

\subsubsection{Smart homes/ Microgrids}

Smart home algorithms often include EVs as one of the demand response appliances that help minimizing the total home electricity bill \cite{Erdinc2015}, \cite{Wi2013}, \cite{Pal2018}, \cite{Rana2018}, \cite{Naghibi2018}, \cite{Melhem2017}, \cite{Kikusato2018}. These algorithms observe only a single EV at a single location, which directly makes them susceptible to \emph{Issues 1 \& 2}. 

EVs and smart homes can also be grouped under a microgrid where EVs act as flexibility providers \cite{Tushar2014}, \cite{Paterakis2016}. In a future interconnected smart grid, EVs will be able to interact with both the smart communities (local microgrids) and the central grid to offer their services \cite{Zhang2016comm}. Table \ref{tab:issues} shows that papers related to home/microgrids are mostly CS-based and focused on home-chargers with fixed power levels (\emph{Issue 3}) and consider only energy prices (\emph{Issue 4}). Exception to the standard CS-based models are papers \cite{Melhem2017} and \cite{Kikusato2018}, which model the EV behavior throughout the day in a parking-driving sequence, but neglect the possibility of charging at other CSs. In addition to home charging, only paper \cite{Zhang2016comm} considers parking lots and charging stations, but as independent entities capable of utilizing the EVs' flexibility.

\subsubsection{EV aggregators}

Apart from observing a single CP or locational aggregation through microgrids, EVs can be seen as a decentralized source scheduled by an aggregator and without considering their location. Such models can have various goals, such as minimizing the EVs' total charging costs \cite{Nguyen2014}, \cite{Wei2016},\cite{Bessa2013},  \cite{Bessa2012}, \cite{Ortega-Vazquez2013},  \cite{Vagropoulos2013}, minimizing frequency deviations \cite{Vagropoulos2015}, \cite{Kaur2019} maximizing conditional value-at-risk \cite{Wu2016}, optimizing reserve provision \cite{Goebel2016} or maximizing revenue \cite{Sarker2016}. Table \ref{tab:issues} shows that papers related to EV aggregators mostly focus on home chargers within the CS-based concept (\emph{Issues 1 and 2}) and consider fixed power levels and only energy prices (\emph{Issues 3 and 4}). For example, paper \cite{Kaur2019} presents a CS-based framework where aggregators group CSs while EVs migrate among them. On the other hand, authors of \cite{Sarker2016} and \cite{Ortega-Vazquez2013} do indeed model EVs' behaviour throughout the day, but only as availability periods at unspecified types of chargers, i.e. they do not address the fact that EVs charge and discharge at other CSs as well.

Although papers \cite{Nguyen2014}, \cite{Wei2016}, \cite{Vagropoulos2015},  \cite{Bessa2013}, \cite{Wu2016}, \cite{Bessa2012}, \cite{Kaur2019}, \cite{Vagropoulos2013}, \cite{Sarker2016}, \cite{Ortega-Vazquez2013} model EVs connected to the distribution grid, they take into account only energy and/or balancing prices without network fees or infrastructure costs (\emph{Issue 4}).

\subsubsection{Parking lots/ Charging stations}

In addition to residential parking, EVs can also be charged at workplace/commercial/leisure parking lots or fast charging stations. EVs park at parking lots for longer times and power capacity of AC CPs is usually low to medium. On the other hand, EVs do not park at fast (DC) charging stations but only stop to charge, resembling the existing gas stations. Both the smart parking lot and charging station algorithms aim either at maximizing the benefits \cite{Zhang2018}, \cite{Zhang2016}, \cite{Awad2017}, \cite{Kuran2015},  \cite{Akhavan-Rezai2016}, \cite{Mehrabi2018} or minimizing electricity costs \cite{Zhang2017}, \cite{Zheng2019}, \cite{Zhou2019}, \cite{You2016}, while preserving customers expectations. Table \ref{tab:issues} shows that papers related to parking lots/charging stations are CS-based and specific locations are observed without proper multiple power levels (\emph{Issue 3}) or costs (\emph{Issue 4}).

Papers \cite{Zhang2018}, \cite{Zhang2016}, \cite{Zhang2017}, \cite{Awad2017},\cite{Kuran2015}, \cite{Akhavan-Rezai2016} model  workplace/commercial parking lots, while \cite{Mehrabi2018} observes residential private and public parking lots. Similarly, all the papers modeling CS operation tackle a specific CS connected to a single point in the grid and managed by a centralized controller \cite{Zheng2019}, \cite{Zhou2019}, \cite{You2016}, inflicting \emph{Issues 1 \& 2}.

With respect to \emph{Issue 3}, i.e. insufficient power constraints, papers \cite{Zhang2018}, \cite{Zhang2016}, \cite{Zhang2017}, \cite{Akhavan-Rezai2016}, \cite{Mehrabi2018} use fixed CP power constraints at a parking lot or a CS without considering the OBC maximum power. In \cite{Kuran2015} the authors use one fixed value for OBC (the one of Nissan Leaf). Only paper \cite{Awad2017} defines both the EV and the CP power limits, but it only considers CPs at their own parking lot. All the papers investigating CSs use only chargers' power limits without mentioning the OBC power levels \cite{Zheng2019}, \cite{Zhou2019}, \cite{You2016}.

Unlike the majority of papers which do not consider any grid fees (\emph{Issue 4}), the cost of charging in \cite{Awad2017} includes both the electricity price and the grid fees, while \cite{You2016} takes into account battery degradation costs. 

%To summarize, the existing literature is almost exclusively focused on the CS-based e-mobility system where each type of CS is observed and optimized independently. Also, some of the crucial technical and economic data for proper EV scheduling is usually omitted, which could steer the results in a wrong direction.

\begin{figure}[!t]
    \centering
    \includegraphics[trim={0.3cm 2cm 11.5cm 2cm}, clip,width=7cm]{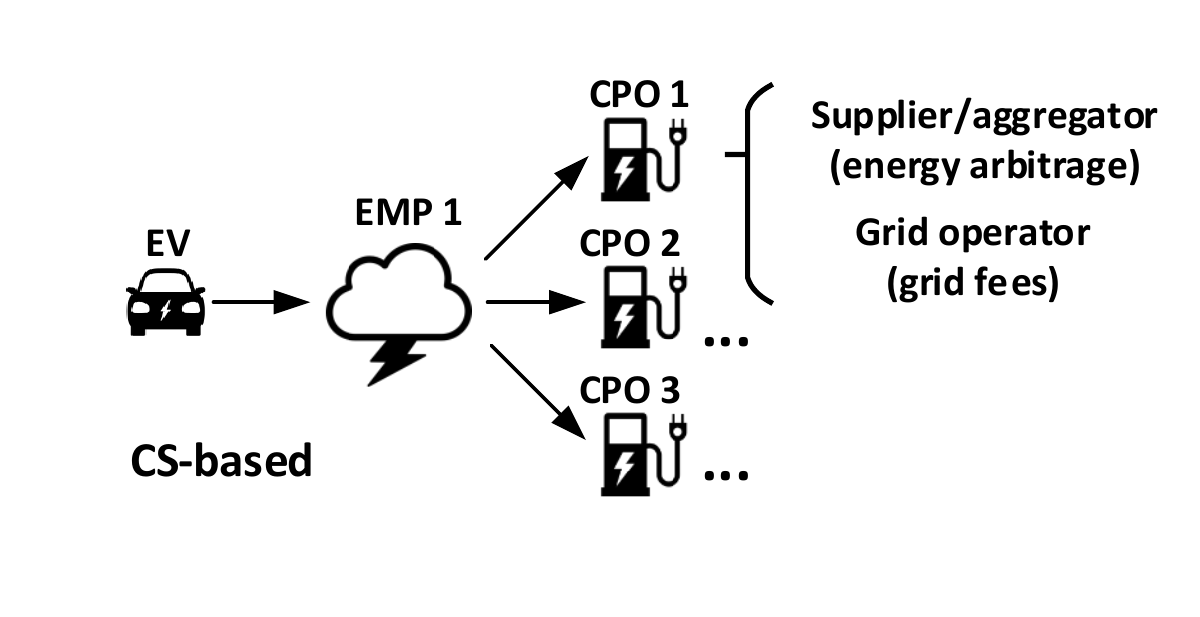}
    \includegraphics[trim={12.0cm 2cm 0.0cm 2cm}, clip,width=7cm]{Ind.pdf}
    \caption{Position of an aggregator in the CS-based and EV-based concepts}
    \label{fig:Ind}
\end{figure}

\subsection{Industry Practices and Research Projects} \label{sec:ind}
 
Current e-mobility related companies can be seen through three business schemes: Charging Point Operators (CPOs), E-Mobility Providers (EMPs), and energy-related companies (electricity suppliers, grid operators). CPOs are the companies operating and maintaining a pool of CPs, while EMPs provide charging services to EV users by enabling them access to CPs (authentication) and offering payment options. EVs have contracts only with EMPs who forward their customers' payments to the CPOs. EMPs have contracts with many CPOs, while the CPOs have contracts with energy suppliers as well as grid operators. If energy arbitrage or flexibility provision through an aggregator is the target, EVs and EMPs cannot directly provide it, only the CPOs can. This is in line with the CS-based smart e-mobility, as illustrated in Figure \ref{fig:Ind}. On the other hand, in the EV-based e-mobility approach the aggregator must be connected to EVs or EMPs. Grid fees are still assigned to CPOs because the physical connection does not change (see Figure \ref{fig:conn}).

The Internet-of-Things (IoT), energy and e-mobility companies already took the CS-based path of the smart e-mobility \cite{Enervalis}, \cite{Wallbox}, \cite{OVOSmartCharger}, \cite{Greenflux2019}. The smart charging in the current industry practices usually means scheduling charging for household users at low electricity tariffs or cutting the peak load of larger CSs. Research projects such as \cite{NewMotion2019}, \cite{SEEV4-City}, \cite{Parker2019} tackle mostly the issue of V2G testing on bidirectional chargers without integrating an aggregator into a real-world e-mobility system.

It is clear that the e-mobility industry does not yet operate within the EV-based smart e-mobility concept, which would change the role of the main beneficiaries in the smart environment from CPOs to EMPs.  

\subsection{Proposed Concept} \label{sec:prop}

The CS-based concept arises from a conventional way of addressing the EVs -- they are an electric load stationary connected at a specific location to a specific CS. This CS does not have information about the EV's battery SOE prior and after the connection and must forecast those values. In this sense, an EVA aggregates specific CSs physically located at households, parking lots or dedicated charging stations and their proper name should be EV Charger Aggregator or EVCA.

We argue that EVs should not be observed as conventional loads but as mobile batteries. EVA should not aggregate specific CSs but the EVs with their batteries. The new concept of EVA is therefore named EV Battery Aggregator or EVBA. EVBA continuously monitors and records EV information (SOE, planed trips) as a part of the future IoT concept. CPOs should allow all EVs to connect without restrictions but for a charging fee. CPOs should be understood as infrastructure operators similar to transmission/distribution system operators and charging a fee in a way that transmission and distribution fees (tariffs) are charged. 

Additional benefits of the EVBA concept are the payment possibilities. Slow chargers are usually part of other consumer facilities and they are controlled within their smart environment (smart households, buildings, parking lots, etc.). It is not quite clear how an EVCA can aggregate CPs at someone else's property. That is why each EV should have its own independent metering device so energy to/from an EV can be exactly measured as in the EVBA case. 

Although EVBA is contrary to scientific research and current industry practises, as discussed in Sections \ref{sec:lit} and \ref{sec:ind}, it is in line with the ISO 15118 standard, which foresees two controllers essential for deployment of a smart e-mobility system: an EV communication controller and a CP communication controller. In such advanced communication architecture, the EVBA can easily communicate the schedules to its EVs and the EVs can send all required data back to the EVBA. The data transfer between EVs and CSs can be easily achieved through EV and CP controllers.     

\section{Conclusion} \label{sec:concl}
This first of the two-paper series on EV smart mobility systems has demonstrated on a small example the shortcomings of the CS-based concept, which is predominantly used in the research community. The main drawback of this concept is that it observes the EV batteries only when connected to a specific CS. This results in suboptimal EV charging schedules and EVCA revenues. Furthermore, CSs have to forecast the EV battery parameters (arrival and departure times and SOE at arrival and departure), which further reduces the optimality of the charging schedule. %However, there is an alternative for forecasting the EV battery parameters, but it comes at a cost of massive communication burden, as well as security issues. 

As opposed to the CS-based concept, which aggregates the CSs, the proposed EV-based concept aggregates EVs themselves. This enables optimal charging schedule for each EV, regardless where it is charged. On top of this, it resolves the communication issues as there is no need for EVs to send their private data to CSs.

Another issue with the current literature is the lack of power constraints. This is related to charging capacities of EV's OBC and CP, as the lower of these two values is binding and, thus, both should be considered in the models. The EVCA concept requires EVs to send the OBC capacity data to CSs in order to determine their future flexibility volume, which is avoided with the EVBA concept.

The final issue we identified are incomplete costs of charging as majority of the published papers do not consider grid fees or infrastructure costs. In the EVCA model, this infrastructure are EVs themselves, which means they should send their costs to CSs so an EVCA can decide on its charging schedule. Again, the proposed EVBA model requires CSs, which are infrastructure in this case, to send their costs to EVs and these costs are already public.

In the second part of this work the EVCA and EVBA concepts are compared on a case study and the impact of each of the four identified issues are quantified.

%paper outlined important findings in the field of smart e-mobility and demonstrated them on the small test case. The commonly observed e-mobility system where the CS takes the leading role at electricity market yields sub-optimal results. The proposed e-mobility system where EV takes the leading role at electricity market proved to be superior from economical perspective. This is especially the case when electricity price volatility is high when EV-based model brings 3,87 times lower costs (for three observed EVs) than in the CS-based case. Opposed to EV-based model, in CS-based models CSs cannot accurately anticipate the optimal arriving and departing SOE and they cannot exchange the flexibility among themselves. 

%Also, the paper showed that insufficiently modeled constraints and costs can steer the scheduling results in the wrong direction leading to infeasible charging/discharging bids and higher real costs then anticipated. Accurate power constraints modeling points out the value of higher installed power both in OBC and external CS equipment. Multiple cost observation can accentuate decision makers which are the areas where they must financially stimulate EVs and their aggregators to attract them to actively participate at electricity markets.

\bibliography{PART_1} 

\end{document}